\documentclass[twocolumn]{autart}    % Enable this line and disable the 
\usepackage[T1]{fontenc}
\usepackage{url}
\usepackage{natbib}
\usepackage{amsmath}
\usepackage{amssymb}
\usepackage{mathtools}

\usepackage{graphicx}          % Include this line if your 
\theoremstyle{plain}
\newtheorem{theorem}{Theorem}[section]

\theoremstyle{definition}

\theoremstyle{remark}
\newtheorem{remark}[theorem]{Remark}

\newcommand{\tr}{\mathrm{tr}}

\DeclareMathOperator*{\argmax}{arg\,max}
\DeclareMathOperator*{\mE}{\mathbb{E}}

\begin{document}

\begin{frontmatter}

\title{Policy Optimization Algorithms in a Unified Framework\thanksref{footnoteinfo}} % Title, preferably not more 

\thanks[footnoteinfo]{
Corresponding author Shuang Wu.}

\author[noahhk]{Shuang Wu}\ead{wushuang.noah@huawei.com}    
\address[noahhk]{Huawei Noah's Ark Lab, Hong Kong SAR, P. R. China} 
          
\begin{keyword}  
policy optimization; ergodicity; reinforcement learning.
\end{keyword}

\begin{abstract}
Policy optimization algorithms are crucial in many fields but challenging to grasp and implement, often due to complex calculations related to Markov decision processes and varying use of discount and average reward setups. This paper presents a unified framework that applies generalized ergodicity theory and perturbation analysis to clarify and enhance the application of these algorithms. Generalized ergodicity theory sheds light on the steady-state behavior of stochastic processes, aiding understanding of both discounted and average rewards. Perturbation analysis provides in-depth insights into the fundamental principles of policy optimization algorithms. We use this framework to identify common implementation errors and demonstrate the correct approaches. Through a case study on Linear Quadratic Regulator problems, we illustrate how slight variations in algorithm design affect implementation outcomes. We aim to make policy optimization algorithms more accessible and reduce their misuse in practice.
\end{abstract}

\end{frontmatter}

\begin{abstract}
Policy optimization algorithms are crucial in many fields but challenging to grasp and implement, often due to complex calculations related to Markov decision processes and varying use of discount and average reward setups. This paper presents a unified framework that applies generalized ergodicity theory and perturbation analysis to clarify and enhance the application of these algorithms. Generalized ergodicity theory sheds light on the steady-state behavior of stochastic processes, aiding understanding of both discounted and average rewards. Perturbation analysis provides in-depth insights into the fundamental principles of policy optimization algorithms. We use this framework to identify common implementation errors and demonstrate the correct approaches. Through a case study on Linear Quadratic Regulator problems, we illustrate how slight variations in algorithm design affect implementation outcomes. We aim to make policy optimization algorithms more accessible and reduce their misuse in practice.
\end{abstract}
\section{Introduction}

Policy optimization algorithms are fundamental in reinforcement learning. Unlike value learning methods, policy optimization algorithms directly learn the optimal policy, mapping states to actions. These algorithms are flexible, applicable to continuous actions and support stochastic policies, making them useful in various fields. Examples include game AI~\citep{berner2019dota}, robotic locomotion~\citep{miki2022learning}, chatbot fine-tuning~\citep{ouyang2022training}, reasoning model training~\citep{deepseekai2025deepseekr1}, robotic manipulation~\citep{ibarz2021train}, character animation~\citep{peng2018deepmimic}, nuclear fusion control~\citep{degrave2022magnetic}, and chip design~\citep{mirhoseini2021graph}.

The main goal of policy optimization algorithms is to search optimal policies for Markov Decision Processes (MDPs). MDPs deal with decision-making under uncertain state changes. Analyzing MDPs is challenging because of the complexity in stochastic processes and the subtleties of discount factors. The discount factors reflect real-world timing preferences and help stabilize learning algorithms but can complicate analysis and interpretation. A notable issue in research is the lack of clear distinction between discounted and average reward setups, which leads to confusion and errors in implementation.

Several studies investigate empirical implementation details and provide guidelines for policy optimization algorithms~\citep{engstrom2019implementation,huang2023implementation}. However, there is a gap in the literature addressing the proper theoretical implementation of these algorithms, with a few exceptions~\citep{thomas2014bias,nota2020policy,wu2022understanding}. \cite{thomas2014bias} pointed out that several natural policy gradient algorithms for discounted objectives yield biased gradient estimates due to ignoring the discount factor. \cite{nota2020policy} highlighted a similar issue in popular policy gradient algorithms, where the update direction is not a gradient of any objective function. \cite{wu2022understanding} demonstrated how to correctly derive policy gradient algorithms in a unified manner for various setups. The common issue relates to appropriately treating the discount factor in implementations. Despite this commonality, it is not obvious how these works are connected or how to systematically avoid potential mistakes when implementing other policy optimization algorithms. There is a need for a comprehensive guide that addresses these topics accessibly for those new to the field, helping demystify the complex terminology.

To properly implement policy optimization algorithms, it is crucial to correctly understand how they work. We propose an easy-to-follow framework that clarifies both the principles of deriving these algorithms and the correct way to implement them. We thus can reduce the chances of making mistakes during the implementation process or, at the very least, help people understand the purpose behind these algorithms better. Achieving this requires a clear and straightforward method for comprehending policy optimization algorithms.

In this paper, we develop a framework to make it easier to understand and connect various policy optimization algorithms. We focus on integrating different setups (discounted, total, and average rewards) and various algorithms, including policy iteration~\citep{howard1960dynamic}, policy gradient~\citep{williams1992simple,sutton1999policy}, natural policy gradient~\citep{kakade2001natural}, trust region policy optimization~\citep{schulman2015trust}, and proximal policy optimization, specifically the PPO-clip version~\citep{schulman2017proximal}.
We use two main tools: a generalized concept of ergodicity and perturbation analysis. \textbf{Ergodicity theory}~\citep{petersen1999ergodic} suggests that over the long term, the time averages of a stochastic process will match up with its space averages, given certain conditions. We expand on this idea of ergodicity to cover both discounted reward and average reward, making it possible to use space-based approaches and simplify the derivation of policy optimization algorithms. \textbf{Perturbation analysis}~\citep{hinch1991perturbation} solves complex problems by starting with a simpler one and adjusting for small changes. This approach is useful for comparing policies and analyzing policy behaviors under perturbations~\citep{cao2007stochastic}. For both discounted and average reward setups, \cite{cao2007stochastic} derived policy iteration and policy gradient algorithms with a primary focus on policy performance comparison. We extend \citep{cao2007stochastic} beyond performance comparison and thus enable the derivation of more policy optimization algorithms such as natural policy gradient, TRPO, and PPO. Furthermore, \cite{cao2007stochastic} used different notations from the contemporary reinforcement learning convention, whereas we adopt a modern notation system that enhances readability.

Our contributions are threefold:
\begin{itemize}
\item We introduce the concept of generalized ergodicity, simplifying the transition from varied complex time-based MDP formulations to one single unified space-based formulations.
\item Based on the space-based formulation, we adopt perturbation analysis~\citep{cao2007stochastic,wu2022understanding} to revisit mainstream policy optimization algorithms and enhance theoretical understanding of these algorithms, which further help us easily extend existing algorithms, e.g., PPO-Clip and TRPO for deterministic policies.
\item With enhance understanding of the policy optimization algorithms, we discuss why existing algorithms are prone to incorrect implementations. We prescribe a guideline for correctly implementing policy optimization algorithms.
\end{itemize}

Our goal is to make the key aspects of different policy optimization algorithms clear to a wide audience. We hope that this paper clarifies their purpose and minimizes errors in implementation caused by misunderstandings. We share our results using a space-based approach and provide practical implementation guides in two tables (Table~\ref{tab: space expression} and ~\ref{tab: implementations}), serving as a convenient reference.

\begin{table}[h]
  \centering
  \caption{Unified expressions in space. We assume that policy $\pi$ is parameterized by $\theta$, i.e., $a\sim\pi_\theta(\cdot|s)$ for stochastic policies and $a=\pi_\theta(s)$ for deterministic policies. We omit the subscript $\theta$ for notation conciseness.}
  \label{tab: space expression}%
  \resizebox{\columnwidth}{!}{
    \begin{tabular}{cc}
    \hline
    quantity      & expression\\
    \hline
    performance metric    &$\displaystyle \int_{s\in\mathbb{S}} \nu_\bullet^{\pi}(ds|s_0) \mE_{a\sim\pi(\cdot|s)}[R(s,a)]$ \\
    performance difference  &$\displaystyle \int_{s\in\mathbb{S}} \nu_\bullet^{\pi'}(ds|s_0) \mE_{a\sim\pi'(\cdot|s)}[A_\bullet^\pi(s,a)]$\\
    stochastic policy gradient & $\displaystyle \int_{s\sim\mathbb{S}} \nu_\bullet^{\pi}(ds|s_0) \mE_{a\sim\pi(\cdot|s)} [\nabla_\theta \log\pi(a|s) Q_\bullet^\pi(s,a)]$\\
    stochastic policy curvature  & $\displaystyle \int_{s\in\mathbb{S}} \nu_\bullet^{\pi'}(ds|s_0) \mE_{a\sim\pi(\cdot|s)} [\nabla_\theta \log\pi(a|s)\nabla_\theta\log\pi(a|s)^\top ]$\\
    deterministic policy gradient & $\displaystyle \int_{s\sim\mathbb{S}} \nu_\bullet^{\pi}(ds|s_0) \nabla_\theta \pi(a) \nabla_a   Q_\bullet^\pi(s,a)$ \\
    deterministic policy curvature 
    & $\displaystyle \int_{s\in\mathbb{S}} \nu_\bullet^{\pi'}(ds|s_0)  [\nabla_{\theta} \pi(s) \nabla_\theta\pi(s)^\top]$\\
    \hline
    \end{tabular}%
  }
\end{table}

\begin{table}[h]
  \centering
  \caption{Empirical estimates of the policy gradient (PG) and policy curvature matrices (PCM) from a single sample path for stochastic (S) and deterministic (D) policies. As the expressions are infinite sums, one needs to truncate them in practice.}
  \label{tab: implementations}
  \resizebox{\columnwidth}{!}{
    \begin{tabular}{ccc}
    \hline
    quantity &discounted ($0<\gamma\leq1$) & average \\
    \hline
    S-PG
    & $\displaystyle\sum_{k=0}^\infty \gamma^k \nabla_\theta \log\pi(a_k|s_k)Q_\gamma^\pi(s_k,a_k) $  
    & $\displaystyle \lim_{T\to\infty}\frac{1}{T+1}\sum_{k=0}^T \nabla_\theta \log\pi(a_k|s_k)Q_\mu^\pi(s_k,a_k)$\\
    D-PG
    & $\displaystyle\sum_{k=0}^\infty \gamma^k \nabla_\theta \pi(s_k) \nabla_a Q_\gamma^\pi(s_k,a_k) $  
    & $\displaystyle \lim_{T\to\infty}\frac{1}{T+1}\sum_{k=0}^T \nabla_\theta \pi(s_k) \nabla_a Q_\mu^\pi(s_k,a_k) $  \\
    S-PCM
    & $\displaystyle\sum_{k=0}^\infty \gamma^k \nabla_\theta \log\pi(a_k|s_k) \nabla_\theta \log\pi(a_k|s_k)^\top $  
    & $\displaystyle \lim_{T\to\infty}\frac{1}{T+1}\sum_{k=0}^T \nabla_\theta \log\pi(a_k|s_k) \nabla_\theta \log\pi(a_k|s_k)^\top$\\
    D-PCM
    & $\displaystyle\sum_{k=0}^\infty \gamma^k \nabla_\theta \pi(s_k) \nabla_\theta \pi(s_k)^\top$  
    & $\displaystyle \lim_{T\to\infty}\frac{1}{T+1}\sum_{k=0}^T \nabla_\theta \pi(s_k) \nabla_\theta \pi(s_k)^\top $  \\
    \hline
    \end{tabular}%
    }
\end{table}
\begin{figure*}[t]
    \centering
    \includegraphics[width=0.9\textwidth]{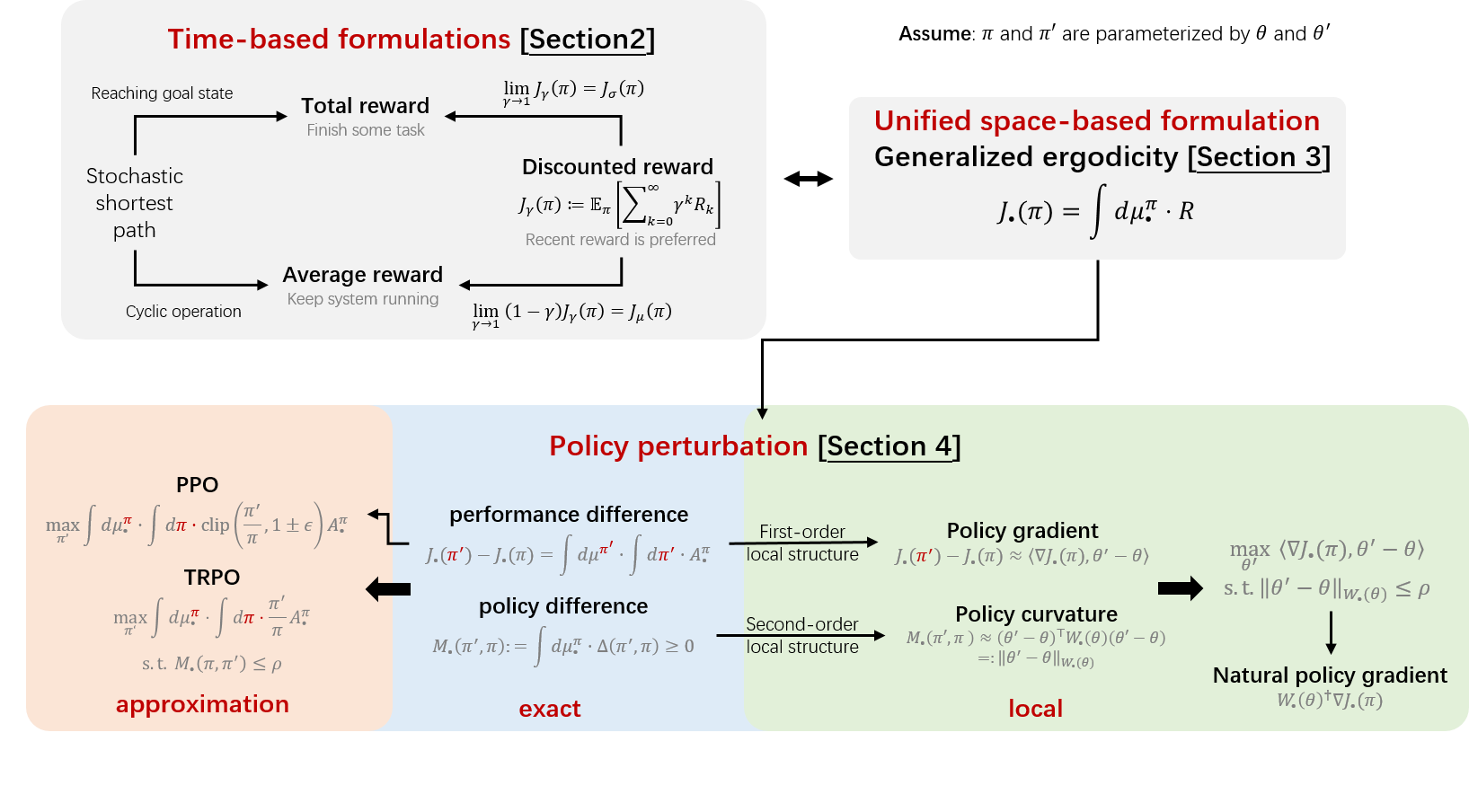}
    \vspace{-3em}
    \caption{Overview of our unified framework. In Section~\ref{section: implementation}, we use these to address incorrect implementations.}
    \label{fig: outline}
\end{figure*}

\section{Problem Setup}\label{section: setup}
This section introduces the Markov Decision Process (MDP) setup. An MDP consists of a state space $\mathbb{S}$, an action space $\mathbb{A}$, a state transition probability $s'\sim\Pr(\cdot|s,a)$, and a reward function $R^\pi(s,a)$, where $s\in\mathbb{S}$ and $s'\in\mathbb{S}$ are states and $a\in\mathbb{A}$ is an action. The process starts with an initial state distribution $s_0\sim\rho_0(\cdot)$. A control policy $a\sim\pi(\cdot|s)$ creates a sequence of states and actions, known as a trajectory $\tau={s_0,a_0,s_1,a_1,\dots}$, where each new state $s_{k+1}$ follows the transition probability $s_{k+1}\sim\Pr(\cdot|s_k,a_k)$.

The expected outcomes of trajectories under a policy $\pi$ are represented as $\mE_\pi[\cdot]$. The policy $\pi$ can be either stochastic (i.e., $a\sim\pi(\cdot|s)$), or deterministic (i.e., $a=\pi(s)$). We assume that the policy $\pi$ is parameterized by $\theta$, allowing us to use $f(\theta)$ and $f(\pi)$ interchangeably. Following~\citep{hernandez1996discrete}, we express both the summation over discrete and integral over continuous variables using the unified form $\int m(dx)f(x) = \sum_x m(x)f(x)$ for discrete $x$, and $\int m(dx)f(x) = \int m(x)f(x)dx$ for continuous $x$.

The general policy optimization problem for MDPs is
\begin{align*}
    \max_\pi \; \mathcal{J}_\bullet(\pi):= \mE_{s_0\sim\rho_0(\cdot)}[J_\bullet(\pi|s_0)],
\end{align*}
which considers three different setups:
\begin{itemize}
    \item \textbf{Discounted} total reward, for $0<\gamma<1$
\begin{align*}
    J_\mathcal{\gamma}(\pi|s_0) = \mE_{\pi} \Big[ \sum_{k=0}^\infty \gamma^k R(s_k,a_k) | s_0 \Big],
\end{align*}
    \item Undiscounted \textbf{total} reward
\begin{align*}
    J_\mathcal{\sigma}(\pi|s_0) = \mE_{\pi} \Big[ \sum_{k=0}^\infty R(s_k,a_k) | s_0  \Big].
\end{align*}
    \item Long-run \textbf{average} reward
\begin{align*}
    J_\mathcal{\mu}(\pi|s_0) = \lim_{T\to\infty}\mE_{\pi} \Big[ \frac{1}{T+1}\sum_{k=0}^T R(s_k,a_k) | s_0 \Big],
\end{align*}
\end{itemize}

Each setup has its critical application scenarios:
\begin{itemize}
    \item \textbf{Discounted}. As the prevalent approach in Reinforcement Learning (RL), it reflects preferences of recent rewards to future rewards and help stabilize learning algorithms. However, the optimal policy depends on the discount factor $\gamma$, and tuning $\gamma$ to achieve a desirable optimal policy often involves trial and error.
    \item \textbf{Total}. This approach is suitable for tasks with a goal state. The MDP should have a terminal state, and at least one policy should lead to this state in a finite expected time. Episodic MDP tasks (finite time) are special cases, incorporating the time step into the state.
    \item \textbf{Average}. This approach aims for sustained high performance and is common in classical dynamic programming, especially in engineering applications with queuing systems \citep{cao1997perturbation}. It's gaining interest in RL, as noted in \cite[Section 10.3 \& 10.4]{sutton2020reinforcement}. For effectiveness, it requires specific properties in the Markov chain, like ergodicity. For broader conditions, such as unichain or multichain, refer to \cite{Kallenberg2002classfication}.
\end{itemize}

Three setups can be linked under specific conditions.
\begin{itemize}
    \item \textbf{Discounted and Total Reward:} Setting $\gamma=1$ in $J^\pi_\gamma$ results in $J^\pi_\sigma$, assuming the total reward is finite (i.e., a terminal state with zero reward is reachable). We will discuss these two setups together. We use the notation for the discounted reward setup, as the results from $J^\pi_\gamma$ can easily apply to $J^\pi_\sigma$ by setting $\gamma=1$.
    \item \textbf{Discounted and Average Reward:} If $J^\pi_\gamma$ becomes unbounded as $\gamma\to1$, indicating cyclic behavior in the MDP, the average reward setup is more suitable. By multiplying $J^\pi_\gamma$ by $1-\gamma$, making $\gamma\to1$ applicable, we derive $\lim_{\gamma\to1} (1-\gamma)J^\pi_\gamma=J^\pi_\mu$, which is the consequence of the Abel theorem~\cite[Lemma 5.3.1]{hernandez1996discrete}. As a side note, the discounted sum is commonly understood as exponential averaging in signal processing, which also indicates that there is a connection between the discounted sum and time average.
    \item \textbf{Total and Average Reward:} In the Stochastic Shortest Path (SSP) framework~(\citealp[Chapter 7]{bertsekas2005dynamic}), the setups are connected. The total reward addresses transient behavior before reaching a goal state. In contrast, the average reward considers the goal state recurrent and seeks the long-term average reward per step. Under mild conditions in SSP, an optimal policy for the undiscounted total reward setup is also optimal for the average reward setup if the total reward MDP has a reset transition that reset the state after reaching goal states.
\end{itemize}

The three time domain setups present challenges in deriving straightforward policy optimization algorithms due to their complex nested structures from stochastic processes and the absence of a standardized approach, potentially leading to misuse of different setups~\citep{thomas2014bias,nota2020policy,wu2022understanding}. In Section~\ref{section: ergodicity}, we introduce the concept of generalized ergodicity to unify these time-based setups into a single state-based framework. Following this, in Section~\ref{section: policy optimization}, we use perturbation analysis to revisit and connect current popular policy optimization algorithms through this unified lens. Figure~\ref{fig: outline} previews our use of generalized ergodicity and perturbation analysis to clarify and link these algorithms. Our goal is to improve understanding and prevent common errors in implementing these algorithms, which is discussed in Section~\ref{section: implementation}.
\section{General Ergodicity}\label{section: ergodicity}
Ordinary ergodicity theory~\cite{petersen1999ergodic} establishes the equality of time averages (averages over sample paths) and space averages (averages over the state space) under suitable conditions (e.g., recurrence). This enables converting results between time-based formulations, useful for Monte Carlo simulations, and space-based formulations, which simplify formula manipulation by dealing with random variables instead of stochastic processes. However, ergodicity is typically associated with the long-run average setup. In this work, we expand this concept to apply comprehensively to both long-run average and discounted setups.

\subsection{Extension and Unification}
Ergodicity theory claims that there exists a probability measure $\nu_\mu^{\pi}(\cdot|s_0)$ equating time and the space averages as below:
\begin{multline}\label{eq: ergodicity average}
    \lim_{T\to\infty}\mE_{\pi} \Big[ \frac{1}{T+1}\sum_{k=0}^T R(s_k,a_k) \mid s_0 \Big] \\= \int_{s\in\mathbb{S}} \nu_\mu^{\pi}(ds|s_0) \int_{a\in\mathbb{A}} \pi(da|s) R(s,a),
\end{multline}
% where $\nu_\mu^{\pi}(\mathbb{X}|s_0) = \lim_{k\to\infty} \mE_{\pi}\Big[\Pr(s_k\in\mathbb{X}|s_0)\Big]$, $\mathbb{X}\subseteq\mathbb{S}$.
where $\nu_\mu^\pi(\mathbb{X}|s_0), \mathbb{X}\subseteq\mathbb{S}$ is the invariant measure of the Markov chain induced by policy $\pi$, i.e., $\nu_\mu^\pi(\mathbb{X}|s_0)=\mE_{a\sim\pi(\cdot|s)}[\int_{s\in\mathbb{S}} \nu_\mu^\pi(ds|s_0) \Pr(\mathbb{X}|s,a)]$. If $\lim_{k\to\infty} \mE_{\pi}[\Pr(s_k\in\mathbb{X}|s_0)]$ exists, then $\nu_\mu^{\pi}(\mathbb{X}|s_0) = \lim_{k\to\infty} \mE_{\pi}[\Pr(s_k\in\mathbb{X}|s_0)]$.

We can extend $\nu_\mu^{\pi}$ to include discounted and total reward cases as
$\nu_\gamma^{\pi}(\mathbb{X}|s_0) = \sum_{k=0}^\infty \gamma^k \mE_{\pi}\Big[\Pr(s_k \in \mathbb{X}|s_0)\Big]$. Eq.~\eqref{eq: ergodicity average} then extends as
\begin{multline}\label{eq: ergodicity discounted}
    \mE_{\pi} \Big[ \sum_{k=0}^\infty \gamma^k R(s_k,a_k) | s_0 \Big] \\ =\int_{s\in\mathbb{S}} \nu_\gamma^{\pi,s_0}(ds|s_0)  \int_{a\in\mathbb{A}} \pi(da|s) R(s,a).
\end{multline}

By combining eqn.~\eqref{eq: ergodicity average} and eqn.~\eqref{eq: ergodicity discounted}, we establish a unified notation for $J_\bullet(\pi|s_0)$ as follows:
\begin{equation}\label{eq: space objective}
    \begin{aligned}
    J_\bullet(\pi|s_0) =&  \int_{s\in\mathbb{S}} \nu_\bullet^{\pi,s_0}(ds)  \int_{a\in\mathbb{A}} \pi(da|s) R(s,a) \\
    =& \int_{s\in\mathbb{S}} \nu_\bullet^{\pi}(ds|s_0) \mE_{a\in\pi(\cdot|s)}[R(s,a)].
\end{aligned}
\end{equation}
We can interpret $\nu_\bullet^\pi$ as an ``averaging" measure. From Table~\ref{tab: space expression}, we can see that the MDP objective, performance difference, policy gradient (stochastic and deterministic), and policy curvature (stochastic and deterministic) are functions of $s$ and $a$ measured by $\nu_\bullet^\pi$. Note that the notation $\bullet$ in $\nu^\pi_\bullet$, $Q^\pi_\bullet(s,a)$ and $A^\pi_\bullet(s,a)$ should consistently used in the same equation. In other words, common expressions such as $\displaystyle \mE_{\substack{s\sim\nu_\mu^{\pi}(\cdot|s_0)\\a\sim\pi(\cdot|s)}} [\nabla_\theta \log\pi(a|s) Q_\gamma^\pi(s,a)]$ for stochastic policy gradient, $\displaystyle \mE_{s\sim\nu_\mu^{\pi}(\cdot|s_0)}[\nabla_\theta \pi(a) \nabla_a  Q_\gamma^\pi(s,a)]$ for deterministic policy gradient, and  $\displaystyle \mE_{\substack{s\sim\nu_\mu^{\pi}(\cdot|s_0)\\a\sim\pi(\cdot|s)}} [\nabla_\theta \log\pi(a|s) \nabla_\theta \log\pi(a|s)^\top]$ for stochastic policy curvatures are not valid as both $\gamma$ and $\mu$ appear in the same expression. We will further address the incorrectness and fixes in Section~\ref{section: implementation}.

\subsection{Application}

\textbf{Time-space conversion}. 
Generalized ergodicity makes it easier to work with different time-based setups by combining them into a single space-based formula using a placeholder $\bullet$. This method is better for deriving formulae because it focuses on random variables instead of stochastic processes. After we come up with a space-based equation, we replace $\bullet$ with specific setups. By looking at the ergodicity equations in eqn.~\eqref{eq: ergodicity average} and~\eqref{eq: ergodicity discounted}, we can transform the equation back to a time-based format. This is useful for Monte Carlo simulations. We will demonstrate how this approach simplifies the development of various policy optimization algorithms in a coherent way and guarantees their correct implementation.

\textbf{Transfer equations between setups}. The generalized ergodicity lets us transfer equations between setups. For example, one found that the correct time-based implementation of policy gradient for the average reward is $\lim_{T\to\infty}\mE_{\pi} \Big[ \frac{1}{T+1}\sum_{k=0}^T \nabla\log\pi(a_k|s_k) Q_\mu^\pi(s_k,a_k) | s_0 \Big]$, then the unified ergodicity leads to following reasoning chain for deriving the counterpart in the discounted setup: 1) space-based gradient (average ergodicity) $\int_{s\in\mathbb{S}} \nu_\mu^{\pi}(ds|s_0)  \int_{a\in\mathbb{A}} \nabla\log\pi(da|s) Q^\pi_\mu(s,a)$, 2) discounted space-based gradient (because of unified form) $\int_{s\in\mathbb{S}} \nu_\gamma^{\pi}(ds|s_0)  \int_{a\in\mathbb{A}} \nabla\log\pi(da|s) Q^\pi_\gamma(s,a)$, and finally 3) discounted time-based gradient (because of time-space conversion) $\mE_{\pi} \Big[ \sum_{k=0}^\infty \gamma^k \nabla \log\pi(a_k|s_k) Q^\pi_\gamma(s_k,a_k) | s_0 \Big]$.
\section{Revisit Policy Optimization}\label{section: policy optimization}
This section builds upon the unified space-based formulation using generalized ergodicity. We introduce a perturbation analysis approach to systematically derive well-known policy optimization algorithms within a unified framework.

\subsection{Performance Difference and Exact Optimization}

In this section, we simplify our notation by omitting $s_0$ from $J_\bullet(\pi|s_0)$ and $\nu_\bullet^\pi(\cdot|s_0)$. This helps avoid confusion without complicating the expressions.

\textbf{Preliminaries}. The state value function and the state-action value function are key for evaluating policy performance. They indicate the future cumulative reward for a current state and state-action pair. For discounted rewards, the state-action value is expressed as
{\allowdisplaybreaks
\begin{align*}
    Q_\gamma^\pi(s, a) =& R(s,a)  +\gamma \int_{s'\in\mathbb{S}} \Pr(ds'|s,a) \\  & \qquad\times \underbrace{\mE_{\pi} \Big[ \sum_{k=0}^\infty \gamma^k R(s_k,a_k)  | s_0=s'  \Big]}_{=V_\gamma^\pi(s')},
\end{align*}
}
while, with average rewards, it takes a different form:
{\allowdisplaybreaks
\begin{align*}
    Q_\mu^\pi(s,a) =& R(s,a) - J_\mu(\pi) +\int_{s'\in\mathbb{S}} \Pr(ds'|s,a) \\ & \qquad\times  \underbrace{\mE_{\pi} \Big[ \sum_{k=0}^\infty R(s_k,a_k) - J_\mu(\pi) | s_0=s'  \Big]}_{=V_\mu^\pi(s')}.
\end{align*}
}
Akin to $J_\bullet$, we adopt a unified notation $Q_\bullet^\pi$ to denote both $Q_\gamma^\pi$ and $Q_\mu^\pi$. This notation allows us to concisely express the value function and advantage function as follows:
\begin{align*}
    V_\bullet^\pi(s) = \mE_{a\sim\pi(\cdot|s)} [Q_\bullet^\pi(s,a)], \;
    A_\bullet^\pi(s,a) = Q_\bullet^\pi(s,a) - V_\bullet^\pi(s).
\end{align*}

\textbf{Performance difference}.
In the context of general ergodicity, the objectives can be expressed as follows:
\begin{align*}
    J_\bullet(\pi)=\int_{s\in\mathbb{S}} \nu_\bullet^\pi(ds) \mE_{a\sim\pi(\cdot|s)}[R(s,a)].
\end{align*}
This equation indicates that the objectives represent the expected single-stage reward under a measure induced by policy $\pi$. The advantage function $A_\bullet^\pi(s,a)$, similar to $R(s,a)$, is a single-stage function. The expression $\mE_{a\sim\pi'(\cdot|s)}[A_\bullet^\pi(s,a)]$ represents the per-stage advantage of policy $\pi'$ over $\pi$. This leads one to guess whether the following \textbf{performance difference formula} is true:
\begin{align}\label{eq: unified space performance difference}
    J_\bullet(\pi')-J_\bullet(\pi)=\int_{s\in\mathbb{S}} \nu_\bullet^{\pi'}(ds)\mE_{a\sim\pi'(\cdot|s)}[A_\bullet^\pi(s,a)].
\end{align}
As stated in~(\citealp[eqn. (20)]{schulman2015trust}), this equation holds in a discounted setup\footnote{We show a state-based proof in the appendix. We can transfer the state-based result to the time-based version using ergodicity.}:
\begin{align}\label{eq: discounted performance difference time}
    J_\gamma(\pi')-J_\gamma(\pi) =  \mE_{\pi'} \Big[ \sum_{k=0}^\infty \gamma^k A_\gamma^\pi(s_k,a_k) \Big].
\end{align}
Drawing on our discussion in the last section, we can generalize the relation to both average and discounted setups by applying generalized ergodicity, thereby verifing eqn.~\eqref{eq: unified space performance difference}. We can also rigorously show why the extension to the average reward setup is correct. By Abel theorem~(\citealp[Lemma 5.3.1]{hernandez1996discrete}), for $0<\gamma<1$,
$\lim_{\gamma \to 1} (1-\gamma) \sum_{k=0}^\infty \gamma^k x_k =
    \lim_{T\to\infty} \frac{1}{T+1} \sum_{k=0}^T x_k$.
By multiplying both sides of eqn.~\eqref{eq: discounted performance difference time} by $(1-\gamma)$ and taking the limit as $\lim_{\gamma\to1}$, we obtain\footnote{We prove $\lim_{\gamma\to1} A_\gamma^\pi(s,a) = A_\mu^\pi(s,a)$ in the appendix.}
\begin{multline}\label{eq: average performance difference time}
    J_\mu(\pi') - J_\mu(\pi)
    = \lim_{T\to\infty}\mE_{\pi'} \Big[ \frac{1}{T+1}\sum_{k=0}^T A_\mu^\pi(s_k,a_k)  \Big].
\end{multline}
We thus rigorously derive the unified performance difference formula in eqn.~\eqref{eq: unified space performance difference}.

\textbf{Policy iteration}.
Since $\nu_\bullet^{\pi'}>0$ for each state, the policy difference formula leads to a simple method to find a better policy, that is, for every $s\in\mathbb{S}$,
\begin{align}\label{eq: policy iteration}
    \pi^{(k+1)}(a|s) = \argmax_{\pi'} \int_{a\in\mathbb{A}} \pi'(da|s) A_\bullet^{\pi^{(k)}}(s,a).
\end{align}
This update is known as policy iteration~\cite{howard1960dynamic}. If $\pi$ is restricted to be a deterministic policy, the policy iteration is simplified to, for every $s\in\mathbb{S}$, $\pi^{(k+1)}(s) = \argmax_{\pi} \; A_\bullet^{\pi^{(k)}}(s,\pi(s))$. The policy iteration scheme is straightforward as one only need to iterate between evaluating $A^\pi(s,a)$ and finding the optimal $a$ of $A^\pi(s,a)$ for each $s$. Since $\nu_\bullet^{\pi'}>0$ for each state, this scheme ensures that the policy sequence from the iteration monotonically attains higher $J_\bullet(\pi)$ values.

\textbf{Policy gradient}. The performance difference formula can be further used to derive a policy gradient formula in a simple way\footnote{\citet{sutton1999policy} developed proofs for stochastic policies in both discounted and average setup, and \citet{silver2014deterministic} showed a proof for deterministic policy with discounted reward only. \citet{wu2022understanding} presented a simpler and more accessible proof using the perturbation method, which we include here for completeness.}. Consider a policy $\pi$ parameterized by $\theta$. According to the definition of gradient, the policy gradient is
\begin{align}\label{eq: policy gradient}
    \nabla_\theta J_\bullet(\theta) = \lim_{\delta\theta \to 0} \frac{J_\bullet(\theta+\delta\theta) - J_\bullet(\theta)}{\delta\theta}.
\end{align}
For \emph{deterministic policies}, we can derive
{\allowdisplaybreaks
\begin{align*}
    \nabla J_\bullet(\theta) 
    =& \lim_{\delta\theta\to0}\int_{s\in\mathbb{S}} \nu_\bullet^{\pi}(ds) \frac{Q^\pi(s,\pi'(s))-Q^\pi(s,\pi(s))}{\delta\theta}\\
    =& \int_{s\in\mathbb{S}} \nu_\bullet^{\pi}(ds) \nabla_\theta Q_\bullet^\pi(s,a=\pi(s))\\
    =& \int_{s\in\mathbb{S}} \nu_\bullet^{\pi}(ds) \nabla_\theta \pi(s) \nabla_a   Q_\bullet^\pi(s,a)|_{a=\pi(s)}.
\end{align*}
}
For \emph{stochastic policies}, we can derive\footnote{As shown in~\citep{wu2022understanding}, the proof can be extended to soft MDPs~\citep{haarnoja2018soft,geist2019theory,mei2020global} where the single stage reward involves an additional entropy-related term $R(s,a)-\tau\log\pi(a|s)$.}
{\allowdisplaybreaks
\begin{align*}
    &\nabla J_\bullet(\theta) \\
    =& \lim_{\delta\theta \to 0} \int_{s\in\mathbb{S}} \nu_\bullet^{\pi'}(ds) \int_{a\in\mathbb{A}} \frac{\pi'(da|s) A_\bullet^\pi(s,a)}{\delta}\\
    =& \lim_{\delta\theta \to 0} \int_{s\in\mathbb{S}} \nu_\bullet^{\pi'}(ds) \int_{a\in\mathbb{A}} \frac{[\pi'(da|s)-\pi(da|s)] Q_\bullet^\pi(s,a)}{\delta\theta}\\
    =& \int_{s\in\mathbb{S}} \nu_\bullet^{\pi}(ds) \int_{a\in\mathbb{A}} \nabla_\theta \pi(da|s) Q_\bullet^\pi(s,a) \\
    =& \int_{s\in\mathbb{S}} \nu_\bullet^{\pi}(ds) \int_{a\in\mathbb{A}} \pi(da|s) \nabla_\theta \log\pi(da|s) Q_\bullet^\pi(s,a).
\end{align*}
}

\subsection{Approximate Policy Optimization}
The performance difference formula requires accesses to $\nu_\bullet^{\pi'}$ instead of $\nu_\bullet^\pi$. The policy iteration greedily finds a better policy by omitting the weighting factor $\nu_\bullet^\pi$. Considering $\nu_\bullet^{\pi'}$ could lead to better policy updates. However, evaluating $\nu_\bullet^{\pi'}$ is costly and even intractable in general. Meanwhile, we have drawn samples from $\nu_\mu^\pi$ to calculate $A_\bullet^\pi(s,a)$\footnote{The discount factor $\gamma$ should be considered for estimating quantities in the discounted setup. Neglecting $\gamma$ leads to incorrect implementations as shown in the next section.}. This motivates us to approximate the performance difference by substituting $\nu_\bullet^{\pi'}$ with $\nu_\bullet^\pi$, which leads to an approximate performance difference
\begin{multline}\label{eq: approximate performance difference}
    J_\bullet(\pi') - J_\bullet(\pi) \approx \\ \int_{s\in\mathbb{S}} \underbrace{\nu_\bullet^{\pi}(ds)}_{\text{not $\nu_\bullet^{\pi'}(ds)$}} \int_{a\in\mathbb{A}} \underbrace{\pi(da|s)}_{\text{not $\pi'(da|s)$}} \frac{\pi'(a|s)}{\pi(a|s)}A_\bullet^\pi(s,a).
\end{multline}
The term $\frac{\pi'(a|s)}{\pi(a|s)}A_\bullet^\pi(s,a)$ is a function of random variables $s$ and $a$, and its ``expectation''\footnote{Note that $\nu_\mu^\pi$ is a probability distribution but $\nu_\gamma^\pi$ is not.} under their joint measure approximates the performance difference. This approximation is only valid when $\pi'$ is close to $\pi$. We discuss two popular approaches for keeping $\pi'$ close to $\pi$.

\textbf{PPO-Clip} from~\citep{schulman2017proximal}. This method discourages $\pi'$ from moving too far away from $\pi$ by clipping the policy ratio as follows
\begin{equation}\label{eq: ppo}
\begin{aligned}
    \frac{\pi'(a|s)}{\pi(a|s)}&A_\bullet^\pi(s,a) \approx \min\Bigg( \; \frac{\pi'(a|s)}{\pi(a|s)}A_\bullet^\pi(s,a), \\
    &\text{clip}\Big(\frac{\pi'(a|s)}{\pi(a|s)}, 1 - \epsilon, 1+\epsilon \Big) A_\bullet^\pi(s,a) \; \Bigg).
\end{aligned}
\end{equation}
While the original PPO-Clip approach is designed for stochastic policies, it faces issues with deterministic policies. In these cases, the ratio $\frac{\pi'(a|s)}{\pi(a|s)}$ becomes problematic. For deterministic policies, the estimated policy difference is better represented as
$J(\pi')-J(\pi)\approx\int_{s\in\mathbb{S}} \nu_\bullet^\pi(ds) A^\pi(s,\pi'(s))$. A suitable PPO-Clip method for deterministic policies can be formulated as
$\max_{\pi'} A_\bullet^\pi(s,\pi'(s)) \; \text{s.t.} \|\pi'(s)-\pi(s)\|\leq \rho$. This deterministic version is more challenging to implement than its stochastic counterpart.

\textbf{Trust region policy optimization} from~\citep{schulman2015trust}. Another approach to constrain $\pi'$ is, while maximizing the approximate performance difference, adding explicit constraints to prevent $\pi'$ from moving too far from $\pi$. In particular, we introduce a metric function $M_\bullet(\pi', \pi):=\int_{s\in\mathbb{S}}\nu_\bullet^{\pi}(ds) \Delta_\bullet^{\pi', \pi}(s)$ to measure the difference between two policies\footnote{The integral over $\nu_\bullet^\pi$ is necessary as it allows a comprehensive evaluation of the policy differences. In particular, if we let $\Delta^{\pi',\pi}=\mE_{a\sim\pi(\cdot|s)}[\frac{\pi'(a|s)}{\pi(a|s)}A^\pi_\bullet(s,a)]$, then $M_\bullet(\pi', \pi)$ will become the approximated performance difference.}. The corresponding constrained optimization is formulated as
\begin{subequations}\label{eq: trpo}
\begin{align}
    \max_{\pi'} \; & \int_{s\in\mathbb{S}}\nu_\bullet^{\pi}(ds) \mE_{a\sim\pi(\cdot|s)}\Big[ \frac{\pi'(a|s)}{\pi(a|s)} A_\bullet^\pi(s,a)\Big]\\
    \text{s.t.} \; & \int_{s\in\mathbb{S}}\nu_\bullet^{\pi}(ds) \Delta_\bullet^{\pi', \pi}(s)=:M(\pi',\pi)\leq \rho.
\end{align}
\end{subequations}
To ensure that $M_\bullet(\pi',\pi)$ is indeed a metric, we require that $\Delta_\bullet(\pi', \pi) \geq 0$ and the equation holds only when $\pi'=\pi$. Furthermore, we require $\Delta_\bullet^{\pi', \pi}(s)$ to be invariant under different parameterization schemes and thus inherent to the policy space. \emph{Stochastic policies} are probability distributions, and the Kullback–Leibler (KL) divergence is a natural choice:
\begin{align*}
    \Delta_\bullet^{\pi', \pi}(s) =  \int_{a\in\mathbb{A}} \pi'(da|s) \log\frac{\pi'(a|s)}{\pi(a|s)}.
\end{align*}
\emph{Deterministic policies} yields actions in a finite-dimensional vector space and a natural choice is the Euclidean distance between control actions (not controller parameters),
\begin{align*}
    \Delta_\bullet^{\pi', \pi}(s) = \|\pi'(s)-\pi(s)\|_2^2.
\end{align*}

\begin{remark}
For two stochastic policies $\pi'$ and $\pi$ characterized by two Gaussian policies parameterized by $\mathcal{N}(\pi'(s),\alpha I)$ and $\mathcal{N}(\pi(s),\alpha I)$, the KL divergence is $\Delta_\bullet^{\pi', \pi}(s)=\frac{1}{2\alpha}\|\pi'(s)-\pi(s)\|_2^2$ (\citealp[p.13]{duchi2007derivations}). As $\alpha\to0$, the policies $\pi'$ and $\pi$ become deterministic policies, and $\lim_{\alpha\to0}\alpha\cdot\Delta_\bullet(\pi', \pi)=\frac{1}{2}\|\pi'(s)-\pi(s)\|_2^2$, which is the policy difference measure for deterministic policies.
\end{remark}

\subsection{Simplified approximation and policy curvature}
Eqn.~\eqref{eq: trpo} is challenging to solve due to the nonlinearity in both the objective and the constraint. We can simplify this by approximating the objective with a linear function in $\theta$, expressed as $(\theta'-\theta)^\top \nabla_\theta J_\bullet(\theta)$. Similarly, we can approximate $\Delta_\bullet^{\pi',\pi}(s)$ within a small region using a quadratic function as
\begin{align*}
    \int_{s\in\mathbb{S}} \nu_\bullet^\pi(ds) \Delta_\bullet^{\pi',\pi}(s)= M_\bullet(\pi',\pi) \approx (\theta'-\theta)W_\bullet(\theta)(\theta'-\theta).
\end{align*}
This approximation scheme is feasible because $\Delta_\bullet(\pi',\pi)\geq 0$ and the equality only holds when $\pi'=\pi$. Using the quadratic approximation, we approximately solve the original trust region optimization in eqn.~\eqref{eq: trpo} as
\begin{subequations}\label{eq: approx trpo}
\begin{align}
    \max_{\theta'} \; & (\theta'-\theta)^\top \nabla_\theta J_\bullet(\theta) \\
    \text{s.t.} \; & (\theta'-\theta)^\top W_\bullet(\theta) (\theta'-\theta) \leq \rho.
\end{align}
\end{subequations}
Solving the KKT conditions in~\eqref{eq: approx trpo} leads to the update rule:
\begin{align*}
    \theta_\mathrm{new} = \theta + \sqrt{\frac{2\rho}{\nabla J_\bullet(\theta)^\top W(\theta)^{\dagger}\nabla J_\bullet(\theta)}}W_\bullet(\theta)^{\dagger}\nabla_\theta J_\bullet(\theta),
\end{align*}
where the superscript $\dagger$ stands for pseudo inverse since $W_\bullet(\theta)$ is not necessarily invertible. This update rule varies the stepsize depending on $\theta$ which requires solving an optimization problem. We can reduce the computation overhead by using a fixed stepsize
\begin{align}\label{eq: npg}
    \theta_\mathrm{new} = \theta + \eta W_\bullet(\theta)^{\dagger}\nabla_\theta J_\bullet(\theta),
\end{align}
where $\eta$ is the stepsize. The update rule eqn.~\eqref{eq: npg} is known as the \textbf{natural policy gradient} algorithm~\cite{kakade2001natural}. Since $M(\pi',\pi)\approx(\theta'-\theta)^\top W_\bullet(\theta)(\theta'-\theta)$, we call $W_\bullet(\theta)$ the \textbf{policy curvature}. It shows the local policy space structure. If $W_\bullet(\theta)\equiv I$, the natural policy gradient simplifies to the policy gradient, overlooking the local structure.

We calculate the policy curvatures for both stochastic and deterministic policies.
For \emph{stochastic policies}, consider $\Delta_\bullet^{\theta',\theta}(s)= \int_{a\in\mathbb{A}} \pi'(da|s) \log\frac{\pi'(a|s)}{\pi(a|s)}$ as a function of $\theta'$. The Taylor expansion of $\Delta_\bullet^{\theta',\theta}(s)$, up to the second order, is\footnote{Note that $\Delta_\bullet^{\theta',\theta}(s) |_{\theta'=\theta}=0$ and $\nabla_{\theta'} \Delta_\bullet^{\theta',\theta}(s) |_{\theta'=\theta}=0$ since $\Delta_\bullet^{\theta',\theta}(s)$ achieves its minimum value $0$ when $\theta'=\theta$.}
$\Delta_\bullet^{\theta', \theta}(s) \approx \frac{1}{2}(\theta'-\theta)^\top H_\bullet(\theta|s)(\theta'-\theta)$. Computing the Hessian matrix $H_\bullet(\theta)$ directly requires back-propagation twice. A more efficient formula is $H_\bullet(\theta|s)  = \mE_{a\sim\pi(\cdot|s)}[\nabla_\theta \log\pi(a|s)\nabla_\theta\log\pi(a|s)^\top ]$. We thus derive
\begin{align*}
    W_\bullet(\theta) =\int_{s\in\mathbb{S}}\nu_\bullet^\pi(ds)\mE_{a\sim\pi(\cdot|s)}[\nabla_\theta \log\pi(a|s)\nabla_\theta\log\pi(a|s)^\top ].
\end{align*}
For \emph{deterministic policies}, the first-order Taylor expansion of $\pi'(s)$ at $\theta$ is
$\pi'(s) \approx \pi'(s)\mid_{\theta'=\theta} + (\theta'-\theta)^\top\nabla_{\theta'} \pi'(s)\mid_{\theta'=\theta} = \pi(s) + (\theta'-\theta)^\top\nabla_{\theta} \pi(s)$.
Therefore, $\pi'(s)-\pi(s)\approx(\theta'-\theta)^\top\nabla_{\theta} \pi(s)$. This allows us to approximate $\| \pi'(s)-\pi(s) \|^2$ as $\| \pi'(s)-\pi(s) \|^2 \approx (\theta'-\theta)^\top [\nabla_{\theta}\pi(s)\nabla_{\theta}\pi(s)^\top] (\theta'-\theta)$. We thus derive
\begin{align*}
    W_\bullet(\theta) =  \int_{s\in\mathbb{S}}\nu_\bullet^\pi(ds) \nabla_{\theta} \pi(s) \nabla_\theta\pi(s)^\top.
\end{align*}

\begin{remark}[TRPO and PPO]
    \citealp{schulman2017proximal} claimed that PPO-Clip is a simplified version of TRPO as it does not require solving the constrained optimization posed by TRPO. Apart from being simpler, PPO also surpasses TRPO in performance, a phenomenon not yet fully understood. Our derivations provide insights into why PPO is more effective. Specifically, PPO-Clip improves policy directly by optimizing the approximated performance difference within a small region.
    In contrast, TRPO practitioners approximate the original, highly nonlinear objective in eqn.~\eqref{eq: trpo} with a linear and a quadratic function for the objective and the constraint, respectively. While these simplifications lead to a more manageable optimization problem, they reduce the effectiveness of the approximation of the approximated performance difference in eqn.~\eqref{eq: approximate performance difference}.
\end{remark}
\section{Issues in Implementation}\label{section: implementation}

Our generalized ergodicity framework prescribes policy optimization algorithms that match $\nu_\bullet^\pi$ with $Q_\bullet^\pi(s,a)$. By applying ergodicity equations, we can transfer space-based equations to time-based equations, which allows us to implement these algorithms correctly for both discounted and average reward setups. Table~\ref{tab: implementations} summarizes the corresponding formulae for implementation.

\subsection{Incorrect Implementations}
A number of previous works~\citep{thomas2014bias,nota2020policy,wu2022understanding} pointed out that policy optimizations are often incorrectly executed, usually in a \textbf{discounted reward} setup.

\textbf{Stochastic policies}.
The typical incorrect policy gradient and policy curvature are:
\begin{align}\label{eq: ambiguous equations}
    \nabla J_\gamma(\pi) =& \mE[\nabla\log\pi(a|s)Q_\gamma^\pi(s,a)],\\
    W_\gamma(\theta) =& \mE[\nabla\log\pi(a|s)\nabla\log\pi(a|s)^\top].
\end{align}
However, these formulae can \textbf{confuse} users into thinking they should calculate expectations under the policy $\pi$'s distribution, leading to \textbf{incorrect} implementations as:
\begin{align*}
    \overset{\text{wrong}}{\nabla} J_\gamma(\pi) = &\sum_{k=0}^\infty  \nabla \log\pi(a_k|s_k) Q_\gamma^\pi(s_k,a_k),\\
   \overset{\text{wrong}}{W}_\gamma(\theta)= &\sum_{k=0}^\infty \nabla \log\pi(a_k|s_k) \nabla \log\pi(a_k|s_k)^\top,
\end{align*}
resulting in biased estimates for $\nabla J_\gamma(\pi)$ and $W_\gamma(\theta)$. In particular, we can show that
{\allowdisplaybreaks
\begin{align*}
        \hat{\nabla} J_\gamma(\theta)
    =& \sum_{k=0}^T \nabla\log\pi(a_k|s_k)Q_\gamma^\pi(s_k,a_k) \\
    \propto& \frac{1}{T}\sum_{k=0}^T \nabla\log\pi(a_k|s_k)Q_\gamma^\pi(s_k,a_k)\\
    \approx& \lim_{T\to\infty}\frac{1}{T}\mE_{\pi}\Big[\sum_{k=0}^T \nabla\log\pi(a_k|s_k)Q_\gamma^\pi(s_k,a_k)\Big]\\
    =& \int_{s\in\mathbb{S}} \nu_\mu^{\pi}(ds) \mE_{a\sim\pi(\cdot|s)} [\nabla_\theta \log\pi(a|s) Q_\gamma^\pi(s,a)]\\
    \neq& \nabla J_\gamma(\theta),
\end{align*}
}
and similarly, for policy curvature
{\allowdisplaybreaks
\begin{align*}
    \hat{W}_\gamma(\theta) 
    =& \sum_{k=0}^T \nabla_\theta\log\pi(a_k|s_k)\nabla_\theta\log\pi(a_k|s_k)^\top \\
    \underset{\sim}{\propto}& \int_{s\in\mathbb{S}} \nu_\mu^{\pi}(ds) \mE_{a\sim\pi(\cdot|s)} \Big[\nabla\log\pi(a|s)\nabla_\theta\log\pi(a|s)^\top \Big] \\
    =& W_\mu(\theta).
\end{align*}
}
Therefore, the incorrect implementations yield a measure over $\nu_\mu^\pi$ instead of $\nu_\gamma^\pi$.

\textbf{Deterministic policies}. The derivation resembles that of the stochastic policies. The typical incorrect policy gradient and policy curvature are:
\begin{align}\label{eq: ambiguous equations for deterministic policies}
    \nabla J_\gamma(\pi) =& \mE[\nabla_\theta\pi(s)\nabla_{a} Q_\gamma^\pi(s,a)],\\
    W_\gamma(\theta) =& \mE[\nabla_\theta\pi(s)\nabla_\theta\pi(s)^\top].
\end{align}
The corresponding \textbf{incorrect} implementations are:
\begin{align*}
    \overset{\text{wrong}}{\nabla} J_\gamma(\pi) = &\sum_{k=0}^\infty  \nabla_\theta\pi(s_k)\nabla_{a} Q_\gamma^\pi(s_k,a)|_{a=a_k},\\
   \overset{\text{wrong}}{W}_\gamma(\theta)= &\sum_{k=0}^\infty \nabla \pi(a_k|s_k) \nabla \pi(a_k|s_k)^\top,
\end{align*}
resulting in biased estimates for $\nabla J_\gamma(\pi)$ and $W_\gamma(\theta)$. We can show that
{\allowdisplaybreaks
\begin{align*}
        \hat{\nabla} J_\gamma(\theta) 
    =& \sum_{k=0}^T \nabla_\theta\pi(s_k)\nabla_{a} Q_\gamma^\pi(s_k,a)|_{a=a_k} \\
    \underset{\sim}{\propto}& \int_{s\in\mathbb{S}} \nu_\mu^{\pi}(ds) \nabla_\theta \pi(s) \nabla_a   Q_\gamma^\pi(s,a)|_{a=\pi(s)}\\
    \neq& \nabla J_\gamma(\theta),
\end{align*}
}
and similarly, for policy curvature
{\allowdisplaybreaks
\begin{align*}
        \hat{W}_\gamma(\theta)
    =& \sum_{k=0}^T \nabla_\theta\pi(s_k)\nabla_\theta\pi(s_k)^\top \\
    \underset{\sim}{\propto}& \int_{s\in\mathbb{S}} \nu_\mu^{\pi}(ds) \nabla_\theta\pi(s) \nabla_\theta\pi(s)^\top\\
    =& W_\mu(\theta).
\end{align*}
}
Therefore, the incorrect implementations yield a measure over $\nu_\mu^\pi$ instead of $\nu_\gamma^\pi$.

\textbf{Summary of incorrect Implementations}
Because of misleading expressions discussed above, a number of incorrect implementations could be produced as listed in Table~\ref{tab: common incorrect implementations}.

\begin{table}[h]
    \centering
    \caption{Examples of common incorrect implementations.}
    \label{tab: common incorrect implementations}
    \resizebox{\columnwidth}{!}{
    \begin{tabular}{ccc}
        \hline
        name & expression & reason \\
        \hline
        hybrid gradient &  $\overset{\text{wrong}}{\nabla} J_\gamma$ & mistake $\nu_\nu^\pi$ as $\nu_\gamma^\pi$\\
        hybrid natural gradient &  $W_\mu^{\dagger}\nabla J_\gamma$ & mistake $W_\mu$ as $W_\gamma$\\
        hybrid natural hybrid gradient &  $ W_\mu^{\dagger} \overset{\text{wrong}}{\nabla} J_\gamma$ & both reasons above\\
        \hline
    \end{tabular}
    }
\end{table}

\subsection{Correct implementations}
In our space-based formulation, for discounted rewards, by substituting $\bullet$ with $\gamma$, the equations for policy gradient and curvature become\footnote{Let $\mE_{\substack{s\sim\nu_\gamma^\pi(\cdot)\\a\sim\pi(\cdot|s)}}[\xi(s,a)] := \int_{s\in\mathbb{S}} \nu_\gamma^\pi(ds)  \int_{a\in\mathbb{A}} \pi(da|s) \xi(s,a)$.}.
\begin{table}[h]
    \centering
    \resizebox{\columnwidth}{!}{
    \begin{tabular}{ccc}
    \hline
        & stochastic policies & deterministic policies \\
    \hline
    $\nabla J_\gamma(\pi)$     
        &  $\mE_{\substack{s\sim\nu_\gamma^\pi(\cdot)\\a\sim\pi(\cdot|s)}}
    [\nabla \log\pi(a|s) Q_\gamma^\pi(s,a)]$
        & $\mE_{\substack{s\sim\nu_\gamma^\pi(\cdot)\\a\sim\pi(\cdot|s)}} [\nabla_\theta\pi(s)\nabla_{a} Q_\gamma^\pi(s,a)]$
    \\
    $W_\gamma(\theta)$
         & $\mE_{\substack{s\sim\nu_\gamma^\pi(\cdot)\\a\sim\pi(\cdot|s)}} [\nabla \log\pi(a|s) \nabla \log\pi(a|s)^\top]$
         & $\mE_{\substack{s\sim\nu_\gamma^\pi(\cdot)\\a\sim\pi(\cdot|s)}} [\nabla \pi(a|s) \nabla \pi(a|s)^\top]$
    \\
    \hline
    \end{tabular}
    }
\end{table}

Using the ergodicity for discounted setup, we derive the correct implementations, which include the discount factor $\gamma^k$ as follows.
\begin{table}[h]
    \centering
    \resizebox{\columnwidth}{!}{
    \begin{tabular}{ccc}
    \hline
        & stochastic policies & deterministic policies \\
    \hline
    $\nabla J_\gamma(\pi)$     
        &  $\sum_{k=0}^\infty \gamma^k  \nabla \log\pi(a_k|s_k) Q_\gamma^\pi(s_k,a_k)$
        & $\sum_{k=0}^\infty \gamma^k \nabla_\theta\pi(s_k)\nabla_{a} Q_\gamma^\pi(s_k,a)|_{a=a_k}$
    \\
    $W_\gamma(\theta)$
         & $\sum_{k=0}^\infty \gamma^k \nabla \log\pi(a_k|s_k) \nabla \log\pi(a_k|s_k)^\top$
         & $\sum_{k=0}^\infty \gamma^k \nabla \pi(a_k|s_k) \nabla \pi(a_k|s_k)^\top$
    \\
    \hline
    \end{tabular}
    }
\end{table}

These expressions align with the discounted total cost objective. In contrast, the incorrect methods directly use sample path sums for estimations, neglecting the discount factor $\gamma^k$ in aggregating quantities along the sample path.

\begin{remark}
    Though incorrect implementations do not necessarily lead to significant errors, recognizing these nuances helps clarify the algorithms' intended functionality and execution.
\end{remark}

\section{Case Studies}
We discuss further how the algorithms are applied using case studies. We use the Linear quadratic regulator (LQR) problem as an example.

\subsection{Linear quadratic regulator}
The LQR problem is fundamental in optimal control theory. The state transition is linear in the state $s\in\mathbb{R}^n$ and the action $a\in\mathbb{R}^m$, expressed as $s_{k+1} \sim \mathcal{N} (\Sigma_A s_k + \Sigma_B a_k + w_k, \Sigma_W)$, where $\Sigma_A\in\mathbb{R}^{n\times n}$ is the system matrix, $\Sigma_B\in\mathbb{R}^{n\times m}$ is the control input matrix, and $w_k\sim\mathcal{N}(0,\Sigma_W)$ is the process noise. The single-stage cost\footnote{We use cost instead of reward to align with the literature, e.g., \citep{bertsekas2005dynamic}. The policy optimization problem minimizes the corresponding aggregated costs.} is a quadratic function, expressed as $C(s,a) = s^\top \Sigma_Q s + a^\top \Sigma_R a$.  We consider all linear controllers $a = \pi(s)=-Ks$, where $K$ is the controller parameter. The associated system dynamics can be simplified to $s_{k+1} \sim \mathcal{N} (\Sigma_{A_K}s_k+ w_k, \Sigma_W)$, where $\Sigma_{A_K} := \Sigma_A-\Sigma_B\cdot K$.

In the discounted cost setup, the value function is quadratic $V_\gamma^{\pi}(s) = s^\top \Sigma_{\gamma,P}^K s$. Here, $\Sigma_{\gamma,P}^K$ is the solution to a Lyapunov equation $\Sigma_{\gamma,P}^K = \Sigma_Q + K^\top \Sigma_R K + \gamma \Sigma_{A_K}^\top \Sigma_{\gamma,P}^K \Sigma_{A_K}$. The state converges to a Gaussian distribution, i.e., $\lim_{k\to\infty} s_{k} \sim \mathcal{N}(0, \Sigma_{\mu,S}^K)$, where $\Sigma_{\mu,S}^K$ solves another Lyapunov equation $\Sigma_{\mu,S}^K = \Sigma_W + \Sigma_{A_K} \Sigma_{\mu,S}^K \Sigma_{A_K}^\top$. For the discounted setup, we define $\Sigma_{\gamma,S}^K$ to be the solution to the following Lyapunov equation $\Sigma_{\gamma,S}^K = \Sigma_W + \gamma\Sigma_{A_K} \Sigma_{\gamma,S}^K \Sigma_{A_K}^\top$. One can verify that $\Sigma_{\mu,P}^K$ and $\Sigma_{\mu,S}^K$ can be calculated by setting $\gamma=1$ in the corresponding Lyapunov equations for $\Sigma_{\gamma,P}^K$ and $\Sigma_{\gamma,S}^K$.

Define $\Sigma_{\bullet,U}^K:=\Sigma_B^\top \Sigma_{\bullet,P}^K \Sigma_B+\Sigma_R$ and $\Sigma_{\bullet,G}^K:=\Sigma_{\bullet,U}^K K - \Sigma_B^\top \Sigma_{\bullet,P}^K \Sigma_A$. Plugging the performance difference equation in eqn.~\eqref{eq: unified space performance difference} into the LQR problem\footnote{Both $\Sigma_{A_K}$ and $\Sigma_{A_{K+\delta}}$ are assumed to have no eigenvalues outside the unit circle on the complex plane to ensure $J(K)$ and $J(K+\delta)$ are bounded.}, we derive
\begin{equation}\label{eq: performance difference for LQR}
\begin{aligned}
    J_\bullet(\underbrace{K+\delta}_{=:K'}) - J_\bullet(K) 
    =& \int \nu_\bullet^{K'}(ds) ( 2 s^\top \delta^\top \Sigma_{\bullet,G}^Ks  + \| \delta s\|^2_{\Sigma_{\bullet,U}^K})  \\
    = & \underbrace{2\tr(\delta^\top \Sigma_{\bullet,G}^K \Sigma_{\bullet,S}^{K'})}_{\text{linear in $\delta$}} + \underbrace{\tr(\delta^\top \Sigma_{\bullet,U}^K\delta \Sigma_{\bullet,S}^{K'})}_{\text{quadratic in $\delta$}}.
\end{aligned}
\end{equation}
The policy gradient is then $\frac{\partial J_K}{\partial K} =2\Sigma_{\bullet,G}^K\Sigma_{\bullet,S}^K$. The policy curvature can be directly computed from $ W_\bullet(K) =  \int_{s\in\mathbb{S}}\nu_\bullet^\pi(ds) \nabla_{K} \pi(s) \nabla_K\pi(s)^\top=\Sigma_{\bullet,S}^K$. The natural policy gradient is then $2\Sigma_{\bullet,G}^K$.

\subsection{Numerical Examples}
We consider an LQR problem as $\Sigma_A=[\begin{smallmatrix}0.9 &0.1\\0 &1.1\end{smallmatrix}]$, $\Sigma_B=[\begin{smallmatrix}0\\1\end{smallmatrix}]$, $\Sigma_W=[\begin{smallmatrix}1 &0\\0 &1\end{smallmatrix}]$, $\Sigma_Q=[\begin{smallmatrix}1 &0\\0 &1\end{smallmatrix}]$, $\Sigma_R=1$, and linear controller parameterized by $K=[\begin{smallmatrix}K_0 &K_1\end{smallmatrix}]$.
% \begin{align*}, 
% \Sigma_A =& \begin{bmatrix}0.9 &0.1\\0 &1.1\end{bmatrix}, 
% \Sigma_B = \begin{bmatrix}0\\1\end{bmatrix}, 
% \Sigma_W = \begin{bmatrix}1 &0\\0 &1\end{bmatrix},\\
% \Sigma_Q =& \begin{bmatrix}1 &0\\0 &1\end{bmatrix},
% \Sigma_R = 1,
% K=\begin{bmatrix} K_0 &K_1\end{bmatrix}.
% \end{align*}
% A linear controller is $K=\begin{bmatrix} K_0 &K_1\end{bmatrix}$.
% $\Sigma_A = [\begin{smallmatrix}0.9 &0.1\\0 &1.1\end{smallmatrix}]$, 
% $\Sigma_B = [\begin{smallmatrix}0\\1\end{smallmatrix}]$, 
% $\Sigma_W = [\begin{smallmatrix}1 &0\\0 &1\end{smallmatrix}]$,
% $\Sigma_Q = [\begin{smallmatrix}1 &0\\0 &1\end{smallmatrix}]$, and $R = 1$.

\begin{figure}[t]
    \centering
    \includegraphics[width=0.5\textwidth]{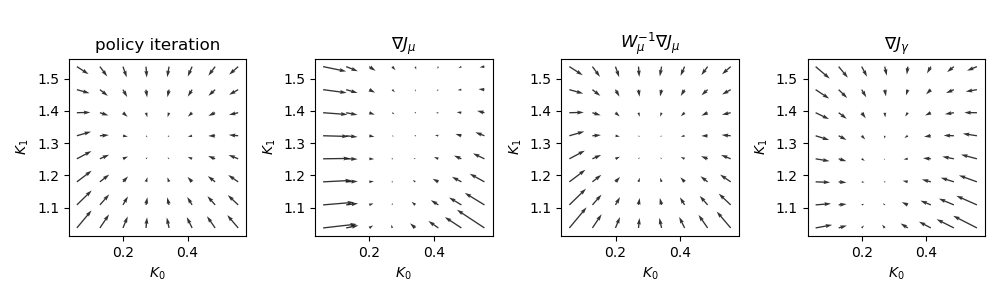}
    \includegraphics[width=0.5\textwidth]{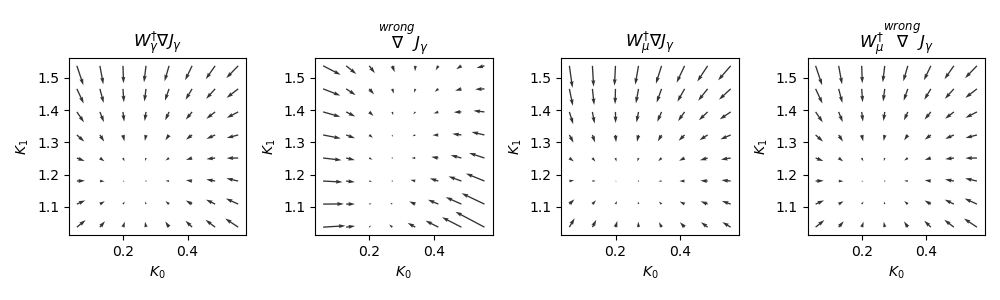}
    \caption{Update directions for both correct and incorrect policy optimization algorithms. Policy iteration, $\nabla J_\mu$, $W_\mu^\dagger \nabla J_\mu$ are correct algorithms for solving $J_\mu$. $\nabla J_\gamma$ and $W_\gamma^\dagger \nabla J_\gamma$ are correct algorithms for solving $J_\gamma$.}
    \label{fig: vector field}
\end{figure}
\begin{figure}[t]
    \centering
    \includegraphics[width=0.45\textwidth]{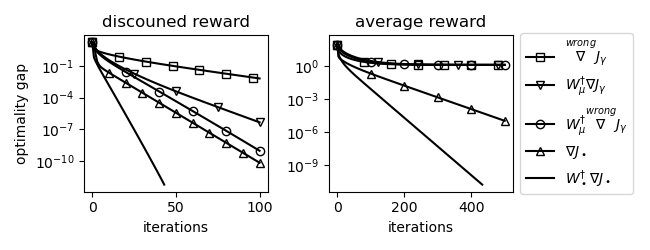}
    \caption{Optimality gap for discounted and average setups.}
    \label{fig: optimality gap}
\end{figure}

\textbf{Vector field of update rules.} We compare the policy update directions of a number of implementations, including policy iteration (for $J_\mu$), policy gradient (for both $J_\mu$ and $J_\gamma$), natural policy gradients (for both $J_\mu$ and $J_\gamma$), and three common incorrect implementations (see Table~\ref{tab: common incorrect implementations} in the appendix for a detailed description).  We chose $\gamma=0.7$ for scenarios involving discounted rewards. 
The generated vector field by these algorithms is depicted in Figure~\ref{fig: vector field}. Note that both the system dynamics and the discount factor $\gamma$ influence the vector field landscape. Additional experiments in the appendix explore the relationship between $\Sigma_A$ and $\gamma$, revealing that slower systems require a larger discount factor to approximate the average reward setup effectively.

\textbf{Optimality gap}. We further investigate how different implementations impact performance, looking at both discounted and average cost setups. This included correct (using $\nabla J_\bullet(\pi)$ and $W(\theta)_\bullet^{\dagger} J_\bullet(\pi)$) and incorrect implementations (the three common ones as just mentioned). 
By using policy iteration as a benchmark for optimal cost, we calculated the optimality gap ($J_\bullet(\pi)-\min_\pi J_\bullet(\pi)$ ) for these methods, shown in Figure~\ref{fig: optimality gap}. We tune the step sizes of different implementations for optimal convergence.  In the discounted cost setup, while incorrect implementations converge, they do so more slowly due to limitations on step sizes. For the average cost setup, incorrect implementations failed to converge to the optimal solution.

\begin{figure}[t]
    \centering
    \includegraphics[width=0.5\textwidth]{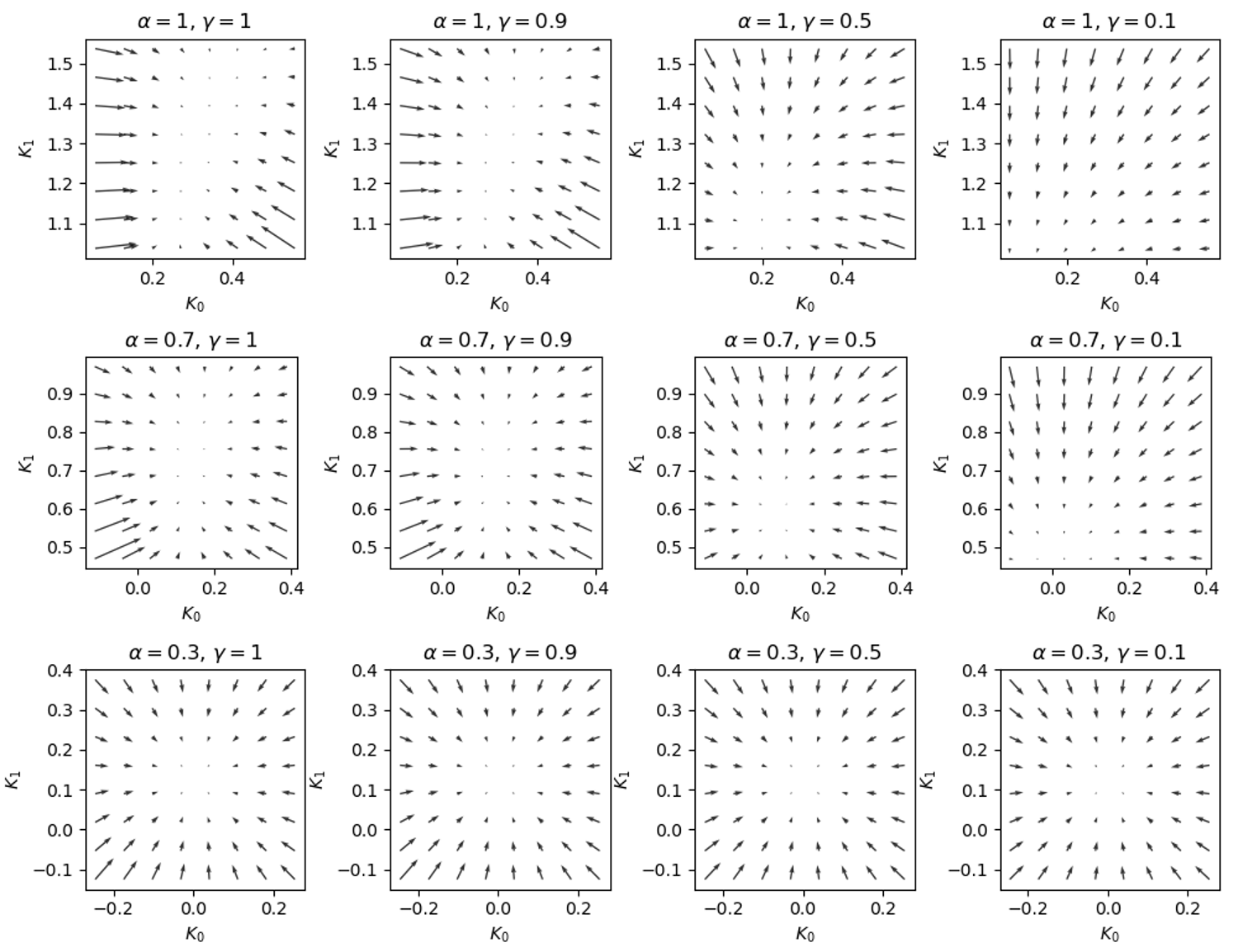}
    \caption{Dynamics ``speed'' and discount factor $\gamma$.}
    \label{fig: dynamics and discount}
\end{figure}

\textbf{Impact of $\gamma$ and MDP structure}. We further investigate how the discount factor $\gamma$ and the convergence speed of the Markov chain affect the vector field. Apart from using a parameterized system dynamics $\alpha[\begin{smallmatrix}0.9 &0.1\\0 &1.1\end{smallmatrix}]$, other system parameters are the same as previous examples. Note that smaller $\alpha$ yields faster system convergence in the absence of control input. Figure~\ref{fig: dynamics and discount} shows the update directions with different system dynamics and discount factors. If the system dynamics converges at a fast rate (small $\alpha$), the difference between using small $\gamma$ and big $\gamma$ is negligible.
\section{Conclusions}
We developed a generalized ergodicity theory to unify different MDP setups in a space-based formulation and leverage perturbation analysis to link existing popular policy optimization algorithms. We hope that this work fosters a comprehensive grasp of policy optimization algorithms and encourages correct implementations.

\bibliographystyle{plainnat} 
\bibliography{ref}           

\appendix
%%%%%%%%%%%%%%%%%%%%%%%%%%%%%%%%%%%%%%%%%%%%%%%%%%%%%%%%%%%%%%%%%%%%%%%%%%%%%%%
%%%%%%%%%%%%%%%%%%%%%%%%%%%%%%%%%%%%%%%%%%%%%%%%%%%%%%%%%%%%%%%%%%%%%%%%%%%%%%%
% APPENDIX
%%%%%%%%%%%%%%%%%%%%%%%%%%%%%%%%%%%%%%%%%%%%%%%%%%%%%%%%%%%%%%%%%%%%%%%%%%%%%%%
%%%%%%%%%%%%%%%%%%%%%%%%%%%%%%%%%%%%%%%%%%%%%%%%%%%%%%%%%%%%%%%%%%%%%%%%%%%%%%%

\section{Proofs}\label{sec: proofs}

\subsection{Proof of $\lim_{\gamma\to1}A_\gamma^\pi=A_\mu^\pi$}
Direct computation yields
{\allowdisplaybreaks
\begin{equation}
\begin{aligned}
    &\lim_{\gamma\to1} A_\gamma^\pi(s_k,a_k) \\
    =& \lim_{\gamma\to1} \Bigg\{ R^\pi(s,a) + \gamma \int_{s\in\mathbb{S}} \Pr(ds'|s,a)V^\pi_\gamma(s') \\ &- R^\pi(s,\pi(s))  - \gamma \int_{s\in\mathbb{S}}  \Pr(ds'|s,\pi(s)) V^\pi_\gamma(s') \Bigg\}\\
    =& R^\pi(s,a) \\
    &+ \int_{s\in\mathbb{S}} \Pr(ds'|s,a) \mE_{\pi} \Big[ \sum_{k=0}^\infty R^\pi(s_k,a_k) | s_0=s' \Big] \\
    &- R^\pi(s,\pi(s)) \\
    &- \int_{s\in\mathbb{S}} \Pr(ds'|s,\pi(s)) \mE_{\pi} \Big[ \sum_{k=0}^\infty R^\pi(s_k,a_k) | s_0=s' \Big] \\
    =& [R^\pi(s,a) - J_\mu(\pi|s_0)] \\
    &+ \int_{s\in\mathbb{S}} \Pr(ds'|s,a) \\
    &\qquad  \times \mE_{\pi} \Big[ \sum_{k=0}^\infty R^\pi(s_k,a_k) - J_\mu(\pi|s_0) | s_0=s' \Big] \\
     & - [R^\pi(s,\pi(s)) - J_\mu(\pi|s_0)]  \\
    & - \int_{s\in\mathbb{S}} \Pr(ds'|s,\pi(s)) \\
    &\qquad \times \mE_{\pi} \Big[ \sum_{k=0}^\infty R^\pi(s_k,a_k) - J_\mu(\pi|s_0) | s_0=s' \Big] \\
    =& [R^\pi(s,a) - J_\mu(\pi|s_0)] + \int_{s\in\mathbb{S}} \Pr(ds'|s,a)V^\pi_\mu(s') \\ &- [R^\pi(s,\pi(s)) - J_\mu(\pi|s_0)] - \int_{s\in\mathbb{S}} \Pr(ds'|s,\pi(s))V^\pi_\mu(s')\\
    =& A_\mu^\pi(s_k,a_k).
\end{aligned}
\end{equation}
}

\subsection{Proof of Performance Difference}
\citet[eqn. (20)]{schulman2015trust} showed that
\begin{align*}
    J_\gamma(\pi'|z)-J_\gamma(\pi|z) =  \mE_{\pi'} \Big[ \sum_{k=0}^\infty \gamma^k A_\gamma^\pi(s_k,a_k) | s_0=z\Big].
\end{align*}
\citet{kakade2002approximately} presented another earlier version of the proof. Both \citet{schulman2015trust} and \citet{kakade2001natural} derived the result in a time-based formulation. We present a proof for the space-based version:
\begin{align*}
    J_\gamma(\pi'|z)-J_\gamma(\pi|z) = \int_{s\in\mathbb{S}} \nu^{\pi',z}_\gamma(ds) \mE_{a\sim\pi'(\cdot|s)}[A^\pi_\gamma(s,a)].
\end{align*}
We consider the vector-based form of Bellman equations for discounted-reward MDPs:
\begin{align*}
    V_\gamma^\pi = R^\pi + \gamma P^\pi V^\pi,
\end{align*}
where $V_\gamma^\pi$ and $R^\pi$ are interpreted as functions of state $s$ and $P^\pi$ is operator on these functions\footnote{If one assumes a finite state space and a finite action space, $V_\gamma^\pi$ and $R_\gamma^\pi$ are vectors, $P^\pi$ is a matrix, and the Bellman equation is a linear equation in $V_\gamma^\pi$.}. We can derive
\begin{align*}
    V^{\pi'}_\gamma - V^\pi_\gamma =& (R^{\pi'} + \gamma P^{\pi'}V_\gamma^{\pi'}) - (R^\pi - \gamma P^{\pi}V_\gamma^{\pi}) \\
    =& (R^{\pi'} + \gamma P^{\pi'}V_\gamma^{\pi}) - (R^\pi - \gamma P^{\pi}V_\gamma^{\pi})\\
    &+ \gamma P^{\pi'}(V_\gamma^{\pi'} - V_\gamma^{\pi}).
\end{align*}
Subtracting $\gamma P^{\pi'}(V_\gamma^{\pi'} - V_\gamma^{\pi})$ on both sides, we obtain
\begin{align*}
    (I-\gamma P^{\pi'})(V^{\pi'}_\gamma - V^\pi_\gamma) = (R^{\pi'} + \gamma P^{\pi'}V_\gamma^{\pi}) - (R^\pi + \gamma P^{\pi}V_\gamma^{\pi}).
\end{align*}
Since $(I-\gamma P^{\pi'})^{-1} = \sum_{k=0}^\infty \gamma^k (P^{\pi'})^k$, we derive 
\begin{align*}
    &J_\gamma(\pi')-J_\gamma(\pi) =V^{\pi'}_\gamma - V^\pi_\gamma \\
    = &\sum_{k=0}^\infty \gamma^k (P^{\pi'})^k [(R^{\pi'} + \gamma P^{\pi'}V_\gamma^{\pi}) - (R^\pi + \gamma P^{\pi}V_\gamma^{\pi})].
\end{align*}
Writing the equation for each component $z$, we derive
\begin{align*}
    J_\gamma(\pi'|z)-J_\gamma(\pi|z) = \int_{s\in\mathbb{S}} \nu^{\pi',z}_\gamma(ds) \mE_{a\sim\pi'(\cdot|s)}[A^\pi_\gamma(s,a)].
\end{align*}
This completes the proof.
%%%%%%%%%%%%%%%%%%%%%%%%%%%%%%%%%%%%%%%%%%%%%%%%%%%%%%%%%%%%%%%%%%%%%%%%%%%%%%%
%%%%%%%%%%%%%%%%%%%%%%%%%%%%%%%%%%%%%%%%%%%%%%%%%%%%%%%%%%%%%%%%%%%%%%%%%%%%%%%

\end{document}